# A new framework for identifying combinatorial regulation of transcription factors: a case study of the yeast cell cycle

Junbai Wang[1] *

1. Department of Biological Sciences, Columbia University, 1212 Amsterdam Avenue, MC 2442, New York, NY 10027, USA

Junbai Wang (**jw2256@columbia.edu**)

**\* Corresponding author**

## Abstract

By integrating heterogeneous functional genomic datasets, we have developed a new framework for detecting combinatorial control of gene expression, which includes estimating transcription factor activities using a singular value decomposition method and reducing high-dimensional input gene space by considering genomic properties of gene clusters. The prediction of cooperative gene regulation is accomplished by either Gaussian Graphical Models or Pairwise Mixed Graphical Models. The proposed framework was tested on yeast cell cycle datasets: (1) 54 known yeast cell cycle genes with 9 cell cycle regulators and (2) 676 putative yeast cell cycle genes with 9 cell cycle regulators. The new framework gave promising results on inferring TF-TF and TF-gene interactions. It also revealed several interesting mechanisms such as negatively correlated protein-protein interactions and low affinity protein-DNA interactions that may be important during the yeast cell cycle. The new framework may easily be extended to study other higher eukaryotes.

## Introduction

### Biological Background of Gene Regulation

A cell can control the proteins it makes by controlling when and how often its genes are transcribed (transcriptional control). For most genes, transcription is controlled by a regulatory region of DNA relatively near the start site of transcription [1]. The regulatory region contains short sequence to which gene regulatory proteins (transcription factors -- TFs) bind. Thousands of TFs and their consensus DNA recognition sequences have been identified. The consensus sequence (TF binding motif) can be used to identify candidate genes whose transcription might be regulated by the TF of interest. However, more direct approaches, such as the chromatin immunoprecipitation technique [2, 3, 4, 5] (i.e. ChIP-chip) can identify TF binding sites in living cells. With the development of other hyphenate high-throughput techniques such as microarray technology for measuring genome-wide expression profiles [1], we are able to study how TFs control genes in response to a variety of signals. In eukaryotic genes, an individual TF can often participate in more than one type of regulatory complex. Such formation of gene regulatory complexes suggests a mechanism for the combinational control of gene expression [6]. In this way, a single gene can respond to an enormous number of combinatorial inputs. Therefore, identifying combinatorial control in eukaryotes is a complex task.

**Computational Approach to the Study of Gene Regulation**

Several methods have been developed to identify combinatorial control of gene expression. For instance, Pilpel et al. [7] looked for cooperatively binding TFs by combining pairs of computationally derived TF binding motifs with gene expression data. In a more recent paper, Yu et al. [8] applied the same motif-based method for identification of interactions between TFs. However, these strategies are hindered because many organisms produce a set of closely related gene regulatory proteins that recognize very similar DNA sequences, and these approaches cannot distinguish between them. To overcome this limitation, Banerjee et al [9] and Kato et al [10] designed algorithms that integrate ChIP-chip data, genome-wide expression, and combinatorial TF-motif analysis to find pairwise TF interactions. Though these integrated approaches increase statistical power for detecting TF interactions, there are two restrictions in their methods, which may bias the outcome in higher eukaryotes: first, the identification of target genes (promoters) or non-target genes of a given TF is only based on ChIP-chip measurement with a manually defined p-value criteria. Such an approach may suffer from the loss of low affinity protein-DNA interactions which may be functionally important [11]. Secondly, the proposed methods are limited to identifying three-way interactions, which excludes high-order phenomena. A number of other recent methods such as Garten et al. [12], Change et al. [13] and Tsai et al. [14] also suffer similar limitations. Although Bar-Joseph et al. [15] described an algorithm that is able to detect high-order TF interactions, their method treats each gene module (gene battery) independently; that is, it does not consider gene-gene interactions when learning the TF-TF interactions. There is another school of research that tries to design mathematical models for inferring transcription factor activities (TFAs), for instance, Li et al. [16], Boulesteix et al. [17], Yang et al. [18] and Kao et al. [19]. Though genome-wide measurement of TFAs remains difficult [20], TFA profiles can be utilized to deduce functional interactions between TFs and to identify putative target genes of transcription factors that are responsible for expression of a gene battery within certain experimental conditions [18]. Therefore, the inferred TFAs are very useful for detecting combinatorial regulation of TFs.

**A New Framework to the Study of Gene Regulation**

In this work, we try to complement the limitations in the early methods and develop a new framework (Figure (1)) for identifying combinatorial regulation of transcription factors. This framework automatically reconstructs a gene regulatory network that includes all possible TF-TF, TF-gene and gene-gene interactions. Our new framework is motivated by initial studies in mathematical modeling of TFA profiles [16, 17, 18, 19] and the reverse engineering of gene regulatory networks from microarray expression data [21, 22]. We first assume that genome-wide expression activities are the product of TFA profiles and TF-DNA affinities (ChIP-chip data) [23]. At the same time, we presume that a cluster of co-expressed genes (gene battery) is controlled by a single TF [1, 24]. For that reason, we can use the singular value decomposition method [25] to compute the TFA profiles. Additionally, we utilize dimensional reduction techniques such as the neural gas algorithm and the stress function [26] to project the high dimensional input gene space onto a low dimensional gene battery space. Subsequently, probabilistic graphical models [27], for example, Gaussian Graphical Models (GGM) or Pairwise Mixed Graphical Models (PMGMS), are applied on the integrated dataset (i.e. TFA profiles and gene battery expression profiles). Thus, by considering both TF activities and gene expression activities in the same uniform framework, all possible TF interactions and TF-Gene interactions can be revealed. In addition, the new framework avoids manual selection of a threshold p-value for identifying protein-DNA binding in the ChIP-chip experiments. Thus, weak TF-DNA interactions may be retained. We suggest several useful features arising from framework: for instance, a new clustering optimization method, visualization of transcription factor activities, and functional enrichment test of MIPS categories [28] for gene batteries. The new features not only simplify the interrogation of complex transcriptional regulations by studying heterogeneous datasets, but also enable us to investigate the detailed mechanism of combinatory regulation of TFs.

To demonstrate that our suggested new framework can be used to identify combinatory regulation of TFs, we tested it in the yeast cell cycle system with gene expression data from Spellman [29] and ChIP-chip occupancy data from Simon [3]. We first applied our framework on 54 yeast genes and 9 transcription factors that are all known to be regulated in the yeast cell cycle. Then, the same approach was used to identify TF-TF and TF-genes interactions in the yeast cell cycle by investigating 676 putative yeast cell cycle genes and 9 TFs. Our results were validated by either genomic sequence data (consensus sequence of protein binding motifs) or literature evidence of cooperativity among transcription factors. At the end of this work, we discuss future improvement of this method.

## Material and Methods

### Sources of experimental data

Microarray experiments of the putative yeast cell cycle regulated genes were obtained from the publication by Spellman et al [29] (676 out of 800, for which DNA sequences of upstream non-coding region are available). There are less than 20% of missing values in the whole dataset. Missing values were imputed by LSimpute [30]. ChIP-chip experiments of nine yeast cell cycle transcriptional regulators were taken from the publication of Simon et al [3]. DNA-binding motifs of nine yeast cell cycle transcriptional regulators were selected from the publications of Yu et al [8] and Banerjee et al [9]. Regulatory Sequence Analysis tools [31] were used to extract upstream DNA sequence and to search for occurrences of protein binding sites (strings) within the upstream region. De novo motif discovery software MotifSampler [32] was used to identify putative binding motif of direct target genes that are regulated by corresponding transcriptional regulators. A full list of 54 known yeast cell cycle genes [3] and their 9 regulators are shown in Table (1).

**Dimensional reduction and Gene battery assignment**

We used a neural gas algorithm [34] to reduce a high-dimensional input gene space $\xi$, ($\xi \in \{g_1, \ldots, g_n\}$) into a low-dimensional "gene battery" space w, ($w \in \{c_1, \ldots, c_C\}$), where reference vector w represents center of each "gene battery". The "gene battery" describes a set of functionally linked genes expressed together for a specific reason that their cis-regulatory systems respond to common trans-regulatory inputs [24, 33]. To estimate the boundary of "gene battery" space w, we applied a modified version of forward search algorithm with stress function [26] to find the best neuron size. To assign each gene into the best class (gene battery), we utilized fuzzy nearest prototype algorithm (FNP) [35] to solve this problem. A detailed description of each method can be found in the web supplement [36].

**Estimating transcription factor activity profiles: Singular Value Decomposition**

Gene expression is controlled by many steps in a cell (for example, transcriptional control, RNA processing control, RNA transport and localization control and protein activity control etc). Here we only focus on genes that are regulated by gene regulatory proteins and the specific DNA sequences (cis-regulatory elements) that these proteins recognize. We assume that the expression of gene batteries with their cis-regulatory systems respond to common trans-regulatory inputs is a linear system [23], which can be written compactly in terms of matrices as follows: the gene expression profile E (n rows (genes) and c columns (experimental conditions)) is obtained from the transcription factors (TFs) activation A (t rows (TFs) and c columns (experimental conditions)) by a linear operation,

$$E=MA \qquad (2)$$

, where M (n rows (genes) and t columns (TFs)) is either the over-representation of cis-regulatory motif or the ChIP-chip measurement of TFs occupancy. The TFs activation matrix A is obtained by operation:

$$A=M^{-1}*E \qquad (3)$$

, where the inverse of matrix M is computed by the singular value decomposition (SVD) method [25]:

$$[u,s,v]=SVD(M), M=u*s*v^T \qquad (4)$$

then

$$A=v*(s^T*s)^{-1}*s^T*u^T*E \qquad (5)$$

In equations (4) and (5), the superscript T means that matrices are transposed.

Generally, transcription factor activation profiles A describe the activity of TFs in a series of experimental conditions. Those conditions are often used by microarray experiments to measure the gene activities E. If we carefully compare the activation patterns of transcription factors and target genes then we can define whether a transcription factor is a relevant regulator of the target [1]. Consequently, we may identify combinatorial regulation of TFs (for example, TF-TF interactions and TF-gene interactions).

**Using continuous transcription factor activity profiles to identify combinatorial regulation of transcription factors: Gaussian Graphical models**

By integrating the estimated activities A of transcription factors and measured expression patterns E of putative target genes into a uniform framework, we can apply reverse engineering techniques such as Bayesian networks, Boolean networks, the S-system, and the probabilistic graphical models [37, 38, 39] to reveal the associations among DNA-binding proteins and their target genes. Here we used Gaussian graphical models with forward search algorithm (GGMF) [22] to identify combinatorial regulation of TFs because there are feedback loops in the yeast cell cycle network. In addition, GGMF is not so sensitive to the rank order of input matrix and the estimated protein activation profiles are continuous. A short description of GGMF will be shown below (please refer to [21, 22] for detailed algorithm): given an independence graph G that is defined by pairwise Markov properties and an input matrix X (for example matrices A and E) with a multivariate normal distribution, we use a covariance selection model [21, 22] to find out the best independence graph consistent with the data, where two variables (gene or TF) are independent given remaining variables when their corresponding element of the partial correlation coefficient matrix is zero [21, 22]. Then, we use an iterative forward search algorithm to search for the potential zero elements in matrix and update partial correlation coefficient matrix with maximum likelihood estimates. The significance level (p-values) of our selection is 0.05, and p-values are adjusted by Bonferroni correction (the normal p-value is multiplied by the number of genes and TFs being tested).

**Using continuous transcription factor activity profiles to find the optimal size of gene batteries**

Usually, it is difficult to identify the optimal subspace from a high dimensional input gene space. Though we can use the stress function and FNP method to estimate the best cluster size, it is better that we apply a method that not only considers the statistical significance of the number of clusters but also dependents on the genomic property of the gene such as mRNA expression pattern, ChIP-chip occupancy data and DNA sequence. Thus, we developed a new clustering optimization method that takes into account the mechanism of control gene expression in a cell: transcription is controlled by short stretches of DNA sequence near the start site of transcription and gene regulatory proteins that recognize and bind to them. These two components operate to turn genes on and off in response to a variety of signals.

Based on the similar consideration, we had demonstrated that there is a linear relationship (E=MA) between the gene expression patterns E (mRNA expression data) and the protein-DNA binding affinity M (ChIP-chip data) in the early section. Now, we make the second assumption that protein binding motifs contribute independently to the binding, such that the total binding affinity (ChIP-chip data M) of protein-DNA interaction is equivalent to the mere sum of the numbers of the individual binding motifs (motif counts C). This additivity assumption may provide a good approximation of true protein-DNA interactions [40, 41]. Thus, we can predict the gene expression activities E' (E'=CA) if transcription factor activity A and its binding motif counts C are known. The predicted gene expression profiles E' may closely resemble the observed gene expression patterns E if the additivity rule of protein-DNA interactions is adequate and the partition of gene batteries is optimal.

Following those assumptions, a new clustering optimization algorithm is described as: start with 2 neurons, during the each iteration, one neuron is added and the FNP is used to assign genes into proper clusters. Then, we compute column sum of ChIP-chip occupancy data M and the corresponding column sum of motif count C in each cluster. The new matrices M' and C' have c rows (neurons) and t columns (TFs). New activity profiles A can be obtained through operation $A = M'^{-1} W$, where matrices W are trained neurons with c rows (neurons) and k columns (conditions). Consequently, we use motif counts C' to predict the expression patterns of gene batteries (neurons) W'=C'A. The error rate of predicted neurons W' is recorded, which is defined as: 1- the number of clusters that have positive correlation coefficients between W and W' and the p-value to their correlations are less than a threshold value (e.g. $p < 0.05$ after correction for multiple testing) divided by the total number of clusters. Optimal neuron size can be determined (minimum error rate) when maximum number of iteration is reached. We tested the new method at a small dataset (54 yeast genes with 9 yeast cell cycle regulators; Table (1)). The test is repeated ten times with 0% replacement of true motif counts C, 30% random replacement of C, 50% random replacement of C and 100% random replacement of C, respectively. The median result of ten tests is recorded.

According to the new searching algorithm, if the number of gene clusters (neurons) reflects the true activities of gene expression patterns then the aggregation of motif counts and ChIP-chip occupancy data in each cluster may work interchangeably through a linear operation. To further test the robustness of such assumption, we tried it at a large dataset (676 putative yeast cell cycle genes with 9 transcription factors). Here we used a slightly modified test procedure: 1) randomly delete 1% of 676 genes; 2) and use the neural gas algorithm to learn the prototypes W of gene batteries; 3) then, we estimate the transcription factor activity profiles A through a linear operation W=MA (M is the ChIP-chip measurements); 4) finally, we predict the prototype W' of gene batteries with the same linear operation W'=C'A but aggregated motif counts C' is used instead of M. The same test is repeated ten times with both genuine protein binding motif counts and randomly sampled motif counts. The error rate of the test is defined the same as the early one. Based on the same test data (676 putative yeast cell cycle genes), the new clustering optimization method was also evaluated against the stress function [26] and the Davies-Bouldin clustering evaluation index [42]. A description of the Davies-Bouldin index can be found in the web supplement [36].

**Using discrete transcription factor activity profiles to identify combinatorial regulation of transcription factors: Pairwise Mixed Graphical models**

In the early section, we proposed a simple method (SVD) to estimate the transcription factor activities (TFAs) A. In literature, there are a number of approaches that can be used to infer the continuous TFAs (for example, the network component analysis [43], the dynamic modeling [44], the partial least square [17] and the regression method [23].) The outcome of these approaches can be directly inputted into our GGMF methods (Figure (1)) for identifying combinatorial regulation of transcription factors. Nevertheless, there are other methods such as state-space model [16] that only provides TFAs with binary results (either on or off). The binary state of TFAs may also be acquired from the prior knowledge of researchers. For instance, all nine yeast cell cycle regulators that were used in this work have well characterized binary TFAs in the literature [3]. Therefore, we developed an alternative model (pairwise mixed graphical models; PMGM) that is able to interrogate a data set with the join distribution of p (TFs) discrete and q (genes or gene batteries) continuous variables. To search for TF-TF and TF-gene interactions, we apply the PMGM on a pair of TFAs with all available gene batteries such search stops when all pairwise TF comparisons are ended.

The mixed graphical models were originally proposed by Lauritzen [45]. In this study, we applied a modified mixed graphical association model -- hierarchical interaction models) [27]. The description of hierarchical interaction models is provided below: we define that the sets of discrete (TFs) and continuous (genes) variables are denoted as $\Delta$ and $\Gamma$ respectively, and a typical observation is written (i,y), where i =(i$_1$, ..., i$_p$) is a p-tuple containing the values of the discrete variables and y is a q vector with real values. Hierarchical interaction models are defined through a parameterization of the condition Gaussian (CG) distribution. This distribution states that the conditional distribution of $\Gamma$ given $\Delta$ is Gaussian (i.e. multivariate normal), and that the marginal distribution of $\Delta$ is arbitrary. Thus the joint density of $\Delta$ and $\Gamma$ is of the form:

$$f(i,y)= p_i(2\pi)^{(-q/2)}|\Sigma_i|^{(-1/2)}\exp\{-1/2(y-u_i)^T\Sigma_i^{-1}(y-u_i)\} \qquad (6)$$

In equation (6) (i,y) belong I (discrete data) and R (continuous data), and p$_i$, u$_i$, $\Sigma_i$ are respectively the probability, mean and covariance of y for 'cell' i (i.e. conditional on $\Delta$=i). Thus {p$_i$} are positive scalar parameters such that $\Sigma_{i\in I}$ p$_i$=1, {u$_i$} are q vectors and {$\Sigma_i$} are positive definite symmetric q x q matrices. To re-parameterize above equation we rewrite it in the form

$$f(i,y)= \exp(\alpha_i + \beta_i^T y - 1/2\ y^T\Omega_i\ y) \qquad (7)$$

where {$\alpha_i$}are scalar parameters, {$\beta_i$}are q vectors and {$\Omega_i$}are positive definite symmetric q x q matrices. We call {$\alpha_i$, $\beta_i$, $\Omega_i$} and {p$_i$, u$_i$, $\Sigma_i$}, the canonical and moment parameterization respectively. The relations between the two parameterizations can be found in the web supplement [36].

Hierarchical interaction models are constructed by restricting the canonical parameters of equation (7) in a similar fashion to hierarchical log-linear [21] models, where the canonical parameters are expanded as sums of interaction terms and models are defined by setting higher-order interaction terms to zero. In other words, if an interaction term is set to zero then all interaction terms that 'include' it are also set to zero. The model parameters subject to these constraints are estimated by a modified iterative proportional scaling algorithm [27]. The significance level (p-values) of model selection is defined the same as Gaussian graphical models.

**A new framework for identifying combinatorial regulation of transcription factors**

In the new framework, we first reduce a high dimensional input gene space (mRNA expression data) into a low dimensional feature space (gene batteries) by applying the Neural Gas, the FNP, and the clustering optimization methods. Here each feature vector (gene battery) represents a cluster of co-expressed genes that may share the same functional category or may be controlled by the same transcription factors in a series of experimental conditions. At the same time, we also apply the SVD method on above input data for estimating the transcription factors activities in conditions relevant to gene expression profiles. We can then integrate both protein activities and gene expression profiles into a uniform dataset because they share the same conditional variation. Consequently, the combined dataset is directly input into reverse engineering algorithms such as the Mixed Graphical Models and the Gaussian Graphical Models for identifying

TF-TF and TF-gene interactions. The workflow of this new framework is shown in Figure (1). It is able to investigate cooperative gene regulation by considering either qualitative or quantitative protein activities.

## Results

**Using continuous transcription factor activity profiles to identify combinatorial regulation of transcription factors: Gaussian Graphical Models with Forward Search Algorithm**

**Defining gene batteries:** From the paper of Simon et al [3], we selected 54 genes that were known to be regulated in the yeast cell cycle. A detailed description of genes and nine transcription regulators are listed in Table (1). We first used the neural gas algorithm and stress function [26] to reduce the high dimensional input gene space into an optimal subspace (11 neurons; please refer to methods section and Figure (1) for detailed description). Then, the fuzzy nearest prototype method [35] was used to find the best gene cluster (battery) for each neuron in which genes were assigned with fuzzy membership values. From this result (Figure (2) and Table (2)) we found that genes with the same functional category are often tightly clustered together such as mitosis control (P=6.30e-05) in cluster 5, mating (P=6.44e-07) in cluster 7, budding (P=3.07e-05, P=3.38e-05) in cluster 8 and 9, cytokinesis (P=1.13e-04) in cluster 10 and chromatin (P=3.00e-09) in cluster 11. However, there are several clusters that contain genes that are involved in diverse functional categories. For instance, genes in each of clusters 1, 2, 4, and 6 belong to at least two functional classes according to the functional categories in Table (1). We also found that several genes having the same biological functions displayed distinct expression patterns and were assigned to different clusters (Table (2)).

For example, both cluster 8 and cluster 9 contain budding genes. Such biases of gene activity may be accounted for by biological phenomena, measurement error, or gene expression noise [46, 47].

**Identifying TF-TF and TF-genes interactions with 54 yeast genes and 9 TFs:** In this work, the singular value decomposition method was used to estimate transcription factor (TF) activities. We expected TF activity profiles to explain the contribution of transcription factors in every experimental condition. Thus, after integrating protein binding activities with gene expression profiles (e.g. Figure (2), prototypes of eleven gene batteries), we applied Gaussian Graphical Models (GGM) on this combined dataset for identifying putative target genes that might be regulated by corresponding transcription factors (TF). A complete transcriptional regulatory network that is predicted by GGM is shown in Figure (3). Detailed description of this network is listed in Table (3a and 3b), where most of the putative TF-gene interactions can be supported by evidence from both literature [3] and DNA sequence (protein binding motif is presented upstream non-coding region) [8, 9]. However, for some putative target genes (such as MCM1-Cluster 3, ACE2-Cluster 4, ACE2-Cluster 8, MBP1-Cluser 5, SWI4-Cluster 7, and SWI6-Cluster 10), we cannot find protein binding motifs of corresponding regulators within the promoter sequences. These TF-gene interactions might be generated by indirect transcriptional regulations (e.g. protein-protein interactions) in the yeast cell cycle because no direct evidence of physical contact between protein and DNA are found. One example is protein NDD1's association with protein SWI6; protein NDD1's putative target gene is cluster 10. Therefore, we could expect an indirect association between SWI6 and cluster 10. For another example, protein ACE2 and protein SWI4 are known to be co-regulated in the yeast cell cycle [3]; protein SWI4 is a part of the DNA binding component of the SBF complex (SWI4-SWI6). In this manner, protein ACE2 could interact with cluster 4 and cluster 9 if these gene batteries are directly regulated by SWI6. In Table (3a), there are a number of indirect protein-DNA interactions such as MCM1-SWI4-Cluster 7-Cluster 3 and MBP1-FKH2-Cluster 5. Those results suggest that protein-protein interactions may play key roles in protein-DNA interactions, such as in the case of two gene regulatory proteins with a weak affinity for each other cooperating to bind to a DNA sequence, neither protein having a sufficient affinity for DNA to efficiently bind to the DNA site on its own. Once two such proteins bind to DNA, the protein dimer creates a distinct surface that is recognized by a third protein carrying an activation domain, stimulating transcription [1, 48]. Results of identifying TF-TF and TF-genes interactions with 676 yeast genes and 9 TFs can be found in the web supplement [36].

**Comparison with previous work for TF-TF interactions among 54 yeast genes and 9 TFs:** Combinatorial regulation by multiple transcription factors is an important problem, and several papers addressing this issue have previously been published before. Three types of methods have been suggested to solve this problem: a method based on co-occurrence of TF binding motifs [8]; a method integrating genome-wide expression data and ChIP-chip data [9]; and the Genetic Regulatory Modules (GRAM) method [15]. In Table (3b), we present a comparison between the TF-TF interactions predicted by our new framework and the TF complexes discovered using above three representative methods, respectively. As can be seen in this table, most of our predicted TF interactions were recovered by the motif-based technique Yu et al. [8], except for Mbp1-Swi5. However, the Mbp1-Swi5 interaction was only found in a genetic regulatory module (Ace2-Mbp1-Ndd1-Swi5) by GRAM [15]. It is clear that the motif-based method cannot identify genetic interactions when there is no physical evidence in the DNA sequence. In the same table, we also found that results by Banerjee et al. [9] are highly dependent on the value of the PB parameter which is a P-value for TF binding to chromatin [2]. For example, with PB<0.0001 (high binding affinity), Banerjee et al. detected only 4 TF-TF interactions, but they recovered 8 interactions when a less stringent 0.01 P-value threshold was used. This tells us that the low affinity protein-DNA interactions may play significant roles in controlling combinatorial gene regulations. Requiring a manually defined P-value (PB) for identifying target or non-target genes of a given TF is a major drawback of the Banerjee et al. approach. In Table (3b), Bar-Joseph et al. [15] identified more transcription factor complexes than Banerjee et al. because GRAM allows the PB cutoff be relaxed if there is sufficient supporting evidence from expression data. However, GRAM missed 4 of our predictions (Ndd1-Fkh1, Mbp1-Fkh1, Swi6-Fkh2 and Swi6-Ndd1) due to a lack of consideration of gene-gene interactions in the Bar-Joseph et al algorithm. Overall, the current comparison (Table (3b)) suggested that the new framework not only overcomes the limitations in the previous methods, but also provides a simple and straightforward way of identifying combinatorial regulation of transcription factors.

**Investigating estimated transcription factor activity profiles:** In Figure (4), we show transcription factor activities of 14 predicted TF-TF interactions. Approximately 70% of these putative TF-TF interactions are negative. Among these negatively associated transcription factors, 70% of them display significant correlation (P<0.05 after correction for multiple testing; for example, NDD1-MCM1 (P=6.0e-03), FKH2-MBP1 (P=3.0e-

05), MCM1-SWI4 (P=1.1e-06), ACE2-SWI4 (P=4.1e-10), FKH2-SWI6 (P=5.0e-05), NDD1-SWI6 (P=5.6e-08) and SWI4-SWI6 (P=2.2e-16). Only a few of the positive correlations were significant such as FKH1-NDD1 (P=1.5e-05) and SWI5-MPB1 (P=9.1e-07). In Figure (4), we also detected several transcription factors such as MCM1 and FKH1 that have much weaker activation profiles than their co-regulators (SWI4 and FKH2) but nevertheless have significant correlations. It seems that the shape of transcription factor activity profiles is more important than the magnitude of TF activities in the combinatorial regulation of transcription. Particularly, negative TF-TF interactions may be the cornerstone for controlling the cooperation of transcription factors in the yeast cell cycle. Such a mechanism would involve a competitive interaction between two gene regulatory proteins, each of which represses the synthesis of the other; this could create flip-flop switch that switches a cell between two alternative patterns of gene expression [1, 49].

**Using direct putative target genes regulated by the corresponding TF to predict protein-binding Motif:** Through our proposed framework in Figure (1), we predicted many putative target genes to be directly regulated by nine yeast cell cycle regulators (detail, Table (3a and 3b)). Based on these putative target genes, protein-binding motifs of nine transcription factors may be recovered from the upstream non-coding region [31, 50]. Thus, to validate our predictions, we extracted 800-bp upstream DNA sequences of the putative target genes from Regulatory Sequence Analysis Tools [31] (RSAT) and applied a motif discovery tool (e.g. MotifSampler [32]) on the DNA sequences. Our predicted DNA-binding motifs of nine yeast cell cycle regulators are listed in Table (4). All putative binding sites closely resemble known motifs, supporting our putative TF-gene interactions. From Table (4), we also observed that one transcription factor usually controls several gene batteries. These gene batteries generally show distinct expression patterns (Figure (2)). This is can be explained by the dynamic nature of TF-DNA interactions. Therefore, putative target genes of the same transcription factor might not have the same expression activities.

To further explore the sequence common to putative target genes, we used RSAT [31] to build feature maps based on known transcription factor binding motifs (Table (4)). Each map represents a DNA sequence located upstream of a given gene and each perfect match of the transcription factor binding site constitutes a feature. A full list of motif occurrences can be found in the web supplement [36]. This analysis shows that not every promoter sequence of the putative target genes contains the DNA-binding motifs of corresponding transcriptional regulators. It further supports our hypothesis that interactions between TFs and genes may be caused by indirect recruitment of transcription factors such as protein-protein interactions.

**Using continuous transcription factor activity profiles to estimate the optimal size of gene batteries:** We first tested our assumption that transcription factor motif count is equivalent to its corresponding ChIP-chip occupancy data, in the new clustering optimization method. We used the neural gas and fuzzy nearest prototype algorithms to classify 54 yeast genes into 11 gene batteries, where the optimal size of gene batteries was decided by the stress function and the prototype W of gene batteries was estimated from raw measurements. Then, we computed a new prototype W' through a linear operation W'=C'A, where C' is the TF motif counts and A is the TF activities. TF activities A were estimated from W=MA, where M is the ChIP-chip measurement. Both W and W' are shown in Figure (5), where the red smooth line represents predicted prototypes W' and the black dashed line represents trained prototypes (neurons) W. From Figure (5), we found that our estimated W' are similar to the trained centers W. P-values to their correlation coefficients are significant (P<0.05 after correction for multiple testing). This result shows that additive TF motif counts of each gene battery can be used interchangeably with their relevant transcription factor occupancy data (ChIP-chip measurement) through a linear operation. Such approximation may achieve the best performance when the optimal subspace of input genes space is found and the estimation of transcription factor activities is reasonable.

To test our new clustering optimization method, we first used the stress function to determine the best subspace (11 neurons) of 54 yeast genes. Then, we applied the new method on the same dataset. The median result of our ten tests is presented in Figure (6) (marked by the blue smoothed line), where neuron size ranges from 8 to 50. In this test, a clear peak of the best predictions appeared when the neuron size was 11. This is in good agreement to the early boundary estimation (marked by the red vertical line in Figure (6)) of the stress function. However, the accuracy of our new method dramatically declined when random noise was introduced into the motif counts C' (red and green lines in Figure (6)). This tells us that the new clustering optimization method only accepts the true sequence properties of the gene batteries. With genuine motif counts, our new optimization method is capable of identifying the optimal feature space from the high dimensional input gene space.

To further test the robustness of new clustering optimization method, we applied it on a large data set (676 potential yeast cell cycle genes and 9 transcription factors). The median performance of ten tests is shown in Figure (7). It is worth noting that the result of genuine protein binding motif counts (blue smoothed line with square) is at least two times better than the outcome of random motif counts (blue smoothed line with cycle). The random motif counts were generated from a normal distribution which has the same minimum and maximum counts as genuine motif counts. In Figure (7), we noticed that stress value continuously decayed as the size of neurons increased from 8 to 100 (black smoothed line with cross). Davies-Bouldin index (normalized by maximum index threshold value 10 and marked by black smoothed line with triangle) has the same decay tendency as the stress value but it starts oscillating frequently after the size of clusters over 60. Such oscillations may indicate the size of neurons over-fits the data. Though it is difficult to estimate the optimal subspace through either the stress function or the Davies-Bouldin index alone, it is easy to identify the best subspace (in this cases 28 neurons, marked by red vertical line in Figure (7)) from 676 yeast genes after we considered the information from the new clustering optimization method. Therefore, by integrating heterogeneous datasets (i.e. the motif count, the ChIP-chip occupancy data, the TF activity and the gene expression data) with a linear system, our new clustering optimization method demonstrated superior results on identifying an optimal low dimensional subspace from high dimensional input gene space when compared to two previous independent approaches.

**Using discrete transcription factor activity profiles to identify combinatorial regulation of transcription factors: Pairwise Mixed Graphical Models**

The model we have described predicts combinatorial regulation of transcription factors in the yeast cell cycle using the GGM method on an integrated dataset such as the combination of transcription factor activities and gene expression profiles. Our results are promising, but GGM models are restricted only to continuous variables. In realistic problems, we may encounter binary transcription factor activities [16]. Therefore, we developed a new pairwise mixed graphical model (PMGM) that considers both continuous and discrete variables in the identification of transcription factors cooperativity. From the data of Simon et al. [3], we collected ON or OFF information of 9 transcription factors in four phases (G1, S, G2 and M) of the yeast cell cycle. Then, we integrate binary protein activities with the same gene expression profiles that we had used previously. Subsequently, PMGM was used to identify putative TF-TF and TF-gene interactions.

In Table (5), we list all of predicted putative TF-TF interactions where the threshold value for the model selection was $P<0.05$ after correction for multiple testing. Among those predicted 22 TF-TF associations, 41 percent of them had been identified by GGM, such as FKH-1-FKH2, FKH2-MBP1, FKH2-NDD1, FKH2-SWI6, NDD1-SWI4, NDD1-SWI6, MBP1-SWI6, MCM1-SWI4 and SWI4-SWI6 (Table (3b)). Among the remaining putative TF-TF interactions (Table (5)), a number were supported by literature evidences [3]; for example, MCM1-FKH2 and NDD1-MCM1 are known to be co-regulated during the G2 phase of the yeast cell cycle. They also contribute to the activation of other co-regulators (i.e. ACE2-SWI5, MCM1-ACE2 and MCM1-SWI5) in the M phase. All three M phase transcription factors (MCM1, ACE2 and SWI5) regulate CLN3, which activates the protein complex SWI4-SWI6 and MBP1-SWI6 during G1 phase. Thus, these three M phase TFs initiate the subsequent co-regulation of MBP1-SWI4, MCM1-MBP1 and MCM1-SWI6. Additionally, SWI4 and MBP1 are active during late G1 and both of them regulate NDD1. Protein FKH2 is bound to SWI4 promoter through-out the cell cycle. Therefore, it is not surprising that MBP1-NDD1 and FKH2-SWI4 are co-regulated during G1 to G2 phase. Protein FKH1is strongly coupled with FKH2 [51] which in turn actives SWI5/ACE2/MCM1 in M/G1 phase. Therefore, FKH1-ACE2, FKH1-SWI5 and FKH1-MCM1 are co-regulated in M phase [3, 52]. For detailed information of gene batteries and putative TF-gene interactions, please refer to the web supplement [36].

From PMGM analysis, we recovered more putative TF-TF interactions than from GGM. It may be that literature information [3] about transcription factor activities provides a better description of protein activities than the linear estimation of corresponding patterns. Though the noise from experimental measurements potentially dilutes the efficiency of GGM, the overall predictions by GGM are comparable to PMGM. As a result, our new PMGM is a useful option for identifying combinatory regulation of transcription factors when one only has discrete information about protein activities [16].

**Discussions and Conclusions**

In this work, we proposed a new framework for identifying combinatorial control by transcription factors. Our suggested approach is applicable to feature selection, protein activity estimation, protein or gene battery activity visualization, and network reconstruction. For feature selection, we have developed a new clustering optimization method to reduce high dimensional input gene space into a low dimensional prototype subspace. Our new optimization method searches for balance between cis-regulatory motif counts and corresponding protein-DNA affinity data through a linear operation, where we assume that the gene activities are controlled by transcription factors. We then optimize gene battery size. The new clustering optimization method was tested on 54 known yeast cell cycle genes and 676 putative yeast cell cycle genes with 9 cell cycle regulators, respectively. Results of this analysis are promising (Figure (6) and Figure (7)). For very high dimensional input gene space (i.e. 676 putative cell cycle genes), our new clustering optimization method perform much better than pure statistical optimization methods such as the stress function and the Davies-Bouldin index.

To estimate transcription factor activation profiles, we utilize the same linear relationship that was used in the feature selection and presume that gene expression profiles are the product of protein activities and protein-DNA interactions. Then, we compute the transcription factor activities through a linear operation by using the experimental measurements of gene expression activities and protein-DNA affinities. In the current framework, we need not select a threshold p-value [2] for filtering weak protein-DNA interactions. We believe that the low-affinity transcription factor-DNA interactions are important in inferring protein activities. This was also suggested in a recent paper [11] that demonstrating abundant low-affinity transcriptional interactions in vivo. Such weak protein-DNA interactions may be important both evolutionarily and functionally. Therefore, by taking into account all possible protein-DNA interactions, we can avoid potential biases that may be generated by manual selection of p-value criteria for identifying transcription factor-DNA interactions.

For the reverse engineering of gene regulatory networks or the identification of cooperation of transcription factors, we used prototypes of gene batteries with estimated transcription factor activation profiles. This combined dataset significantly simplifies the computation of gene networks through probabilistic graphical models such as GGM and PMGM, because the number of unknown parameters is largely reduced. In addition, we can investigate the mechanism of gene expression activities within each gene battery and the protein activities of pairwise TF-TF interactions (Figure (2), Figure (4) and the web supplement [36]. To verify the putative target genes of each gene battery, we applied the hypergeometric test and MotifSampler program to evaluate functional enrichment and to discover putative protein binding motifs of corresponding regulators, respectively (Table (2) and Table (4)). Our method not only predicts pairwise protein-protein interactions but also suggests higher order protein-protein interactions that contribute to the complexity of gene regulation. In addition, at each step of the workflow, the new framework is able to accommodate alternative methods to the ones proposed; for example, another protein activity estimation method or dimensional reduction technique could be substituted.

Nevertheless, there are three limitations in the current framework. First, the new clustering optimization method requires prior knowledge of protein binding motifs and such information is not always available. However, databases [53, 54] and computational tools [55, 11] can provide sufficient putative protein binding sites in model organisms. Second, the prediction of TF-TF and TF-gene interactions is based on static probabilistic graphical models (i.e. GGM and PMGM) that identify interactions among proteins and genes but do not predict when or how such interactions happen. Such information will definitely enhance the interpretation of complex gene regulatory networks. Therefore, we are going to extend our current static graphical models to dynamical graphical models in the future. Finally, though we obtained promising results after testing the new framework on the yeast cell cycle data, our present framework does not guarantee to provide the same good results for higher eukaryotes with more complex genomes such as Caenorhabditis elegans and Drosophila melanogaster. We plan to refine our proposed framework to meet the challenge of other higher eukaryote systems.

In conclusion, we have suggested a new framework for detecting combinatorial regulation of transcription factors. This framework is capable of reconstructing gene regulatory networks by including both continuous and discrete variables. In addition, we have developed several external features such as a new clustering optimization method, transcription factor activity analysis and functional enrichment test of MIPS categories, to assist the integration of heterogeneous data for interrogating gene networks. The proposed framework was tested successfully on yeast cell cycle data, and revealed many known TF-TF and TF-gene interactions. Particularly, we discovered several interesting network features: for example, there are large negatively correlated protein-protein interactions in the yeast cell cycle; protein-protein interactions may play key roles in protein-DNA interactions; low affinity protein-DNA interactions my be important in controlling combinatorial gene regulations; and gene expression with spiky oscillations may make genes very sensitive to the cell cycle system and respond differently in spite of being controlled by the same transcription factor (please refer to the web supplement [36] for detailed description.) A future development of our approach will be to design a dynamic probabilistic graphical model to investigate transcriptional networks in higher eukaryote systems, where the model will show that at which times and under what conditions the protein-DNA interactions are triggered.

## Acknowledgement

We would like to thank H. Bussemaker for the useful discussion, D. Edwards and S.L. Lauritzen for the helpful suggestions during the implementation of mixed graphical models, L.D. Ward for critical reading of manuscript, and T.A. Ruden for accessing high performance computing resources at the University of Oslo. We also thank the two anonymous reviewers for their help comments.

# Reference

[1]     Alberts B, Johnson A, Lewis J, Raff M, Roberts K, Walter P.: *Molecular Biology of the Cell*. Garland Science, New York and London; 2002.

[2]     Lee TI, Rinaldi NJ, Robert F, Odom DT, Bar-Joseph Z, Gerber GK, Hannett NM, Harbison CT, Thompson CM, Simon I, Zeitlinger J, Jennings EG, Murray HL, Gordon DB, Ren B, Wyrick JJ, Tagne JB, Volkert TL, Fraenkel E, Gifford DK, Young RA.: **Transcriptional regulatory networks in Saccharomyces cerevisiae.** Science 2002, **298**:799-804.

[3]     Simon I, Barnett J, Hannett N, Harbison CT, Rinaldi NJ, Volkert TL, Wyrick JJ, Zeitlinger J, Gifford DK, Jaakkola TS, Young RA.: **Serial regulation of transcriptional regulators in the yeast cell cycle.** Cell 2001, **106:**697-708.

[4]     Harbison CT, Gordon DB, Lee TI, Rinaldi NJ, Macisaac KD, Danford TW, Hannett NM, Tagne JB, Reynolds DB, Yoo J, Jennings EG, Zeitlinger J, Pokholok DK, Kellis M, Rolfe PA, Takusagawa KT, Lander ES, Gifford DK, Fraenkel E, Young RA.: **Transcriptional regulatory code of a eukaryotic genome**. Nature 2004, **431**:99-104.

[5]     Gibbons FD, Proft M, Struhl K, Roth FP.**: Chipper: discovering transcription-factor targets from chromatin immunoprecipitation microarrays using variance stabilization.**  Genome Biol 2005, **6**:R96.

[6]     Smale ST.: **Core promoters: active contributors to combinatorial gene regulation.**  Genes Dev 2001, **15**:2503-8.

[7]     Pilpel Y, Sudarsanam P, Church GM.: **Identifying regulatory networks by combinatorial analysis of promoter elements.** Nat Genet 2001, **29**:153-9.

[8]     Yu X, Lin J, Masuda T, Esumi N, Zack DJ, Qian J.: **Genome-wide prediction and characterization of interactions between transcription factors in Saccharomyces cerevisiae.**  Nucleic Acids Res 2006, **34**:917-27.

[9]     Banerjee N, Zhang MQ.: **Identifying cooperativity among transcription factors controlling the cell cycle in yeast.**  Nucleic Acids Res 2003, **31**:7024-31.

[10]    Kato M, Hata N, Banerjee N, Futcher B, Zhang MQ.**: Identifying combinatorial regulation of transcription factors and binding motifs.** Genome Biol 2004, **5**:R56.

[11]    Tanay A.: **Extensive low-affinity transcriptional interactions in the yeast genome.**  Genome Res 2006, **16**:962-72.

[12]    Garten Y, Kaplan S, Pilpel Y.: **Extraction of transcription regulatory signals from genome-wide DNA-protein interaction data.** Nucleic Acids Res 2005, **33**:605-15.

[13]    Chang YH, Wang YC, Chen BS.: **Identification of transcription factor cooperativity via stochastic system model.** Bioinformatics 2006, in press.

[14]    Tsai HK, Lu HH, Li WH.: **Statistical methods for identifying yeast cell cycle transcription factors.** Proc Natl Acad Sci USA 2005, **102**:13532-7.

[15]    Bar-Joseph Z, Gerber GK, Lee TI, Rinaldi NJ, Yoo JY, Robert F, Gordon DB, Fraenkel E, Jaakkola TS, Young RA, Gifford DK.: **Computational discovery of gene modules and regulatory networks.** Nat Biotechnol 2003, **21**:1337-42.

[16]    Li Z, Shaw SM, Yedwabnick MJ, Chan C.: **Using a state-space model with hidden variables to infer transcription factor activities.** Bioinformatics 2006, **22**:747-54.

[17]    Boulesteix AL, Strimmer K.: **Predicting transcription factor activities from combined analysis of microarray and ChIP data: a partial least squares approach.** Theor Biol Med Model 2005, **2**:23.

[18]    Yang YL, Suen J, Brynildsen MP, Galbraith SJ, Liao JC.: **Inferring yeast cell cycle regulators and interactions using transcription factor activities.** BMC Genomics 2005, **6**:90.

[19]    Kao KC, Yang YL, Boscolo R, Sabatti C, Roychowdhury V, and Liao JC.:

**Transcriptome-based determination of multiple transcription regulator activities in

Escherichia coli by using network component analysis**. Proc Natl Acad Sci USA 2004,

**101**:641-646.

[20]    Hoeffler W.: **Method for determining transcription factor activity and its

technical uses.** United States Patent 6913880

[http://www.freepatentsonline.com/6913880.html]

[21]    Wang J, Myklebost O, Hovig E.: **MGraph: graphical models for microarray

data analysis.** Bioinformatics 2003, **19**:2210-1.

[22]    Wang J, Cheung LW, Delabie J.: **New probabilistic graphical models for

genetic regulatory networks studies.**  J Biomed Inform 2005, **38**:443-55.

[23]    Gao F, Foat BC, Bussemaker HJ.: **Defining transcriptional networks through

integrative modeling of mRNA expression and transcription factor binding data.**

BMC Bioinformatics 2004, **5**:31.

[24]    Sandelin A, Hoglund A, Lenhard B, Wasserman WW.: **Integrated analysis of

yeast regulatory sequences for biologically linked clusters of genes.** Funct Integr

Genomics 2003 **3**:125-34.

[25]     Wall ME, Rechtsteiner A, Rocha LM.: *Singular value decomposition and principal component analysis. in A Practical Approach to Microarray Data Analysis.* Kluwer, Norwell, MA; 2003.

[26]     Wang J, Bo TH, Jonassen I, Myklebost O, Hovig E.: **Tumor classification and marker gene prediction by feature selection and fuzzy c-means clustering using microarray data.** BMC Bioinformatics 2003 **4**:60.

[27]     D. Edwards.: *Introduction to graphical modeling.* Springer, New York; 1995.

[28]     Mewes HW, Albermann K, Heumann K, Liebl S, Pfeiffer F.: **MIPS: a database for protein sequences, homology data and yeast genome information.** Nucleic Acids Res 1997, **25**:28-30.

[29]     Spellman PT, Sherlock G, Zhang MQ, Iyer VR, Anders K, Eisen MB, Brown PO, Botstein D, Futcher B.: **Comprehensive identification of cell cycle-regulated genes of the yeast Saccharomyces cerevisiae by microarray hybridization.** Mol Biol Cell 1998, **9**:3273-97.

[30]     Bo TH, Dysvik B, Jonassen I.: **LSimpute: accurate estimation of missing values in microarray data with least squares methods.** Nucleic Acids Res 2004, **32**:e34.

[31]    van Helden J.: **Regulatory sequence analysis tools.** Nucleic Acids Res 2003, **31:**3593-6.

[32]    Thijs G., Marchal K., Lescot M., Rombauts S., De Moor B., Rouzé P., Moreau Y.: **A Gibbs Sampling method to detect over-represented motifs in upstream regions of coexpressed genes.** Journal of Computational Biology 2001, **9:**447-464.

[33]    E.H. Davidson.: *Gene Regulatory Systems. Development and Evolution.* Academic press, San Diego; 2001.

[34]    Martinetz TM, Berkovich SG and Schulten KJ.: **Neural-gas network for vector quantization and its application to time-series prediction.** IEEE TNN 1993, **4:**558-569.

[35]    Keller JM, Gray MR and Givens JE Jr.: **A fuzzy k-nearest neighbour algorithm.** IEEE SMC 1985, **15**:580-585.

[36]    Wang JB.: Supplementary information for "A new framework for identifying combinatorial regulation of transcription factors: a case study of the yeast cell cycle". [http://www.columbia.edu/~jw2256/gtarget/index.html] 2006.

[37]    Tavazoie S, Hughes JD, Campbell MJ, Cho RJ, Church GM.: **Systematic determination of genetic network architecture.** Nat Genet 1999, **22**:281-5.

[38]    Beer MA, Tavazoie S.: **Predicting gene expression from sequence.** Cell 2004, **117**:185-98.

[39]    Chua G, Robinson MD, Morris Q, Hughes TR.: **Transcriptional networks: reverse-engineering gene regulation on a global scale.** Curr Opin Microbiol 2004, **7**:638-46.

[40]    Benos PV, Bulyk ML, Stormo GD.: **Additivity in protein-DNA interactions: how good an approximation is it?** Nucleic Acids Res 2002, **30**:4442-51.

[41]    Veitia RA.: **A sigmoidal transcriptional response: cooperativity, synergy and dosage effects.** Biol Rev Camb Philos Soc 2003, **78:**149-70.

[42]    Vesanto J.: **SOM-Based data visualization methods.** Intelligent Data Analysis journal 1999.

[43]    Liao JC, Boscolo R, Yang YL, Tran LM, Sabatti C, and Roychowdhury VP.: **Network component analysis: Reconstruction of regulatory signals in biological systems.** Proc Natl Acad Sci USA 2003, **100:**15522-15527.

[44]    Lin LH, Lee HC, Li WH, Chen BS.: **Dynamic modeling of cis-regulatory circuits and gene expression prediction via cross-gene identification.** BMC Bioinformatics 2005, **6:**258.

[45]     S. L. Lauritzen.: *Graphical Models.* Clarendon Press, Oxford; 1996.

[46]     de Lichtenberg U, Jensen TS, Jensen LJ, Brunak S.: **Protein feature based identification of cell cycle regulated proteins in yeast.** J Mol Biol 2003, **329**:663-74.

[47]     Raser JM, O'Shea EK.: **Noise in gene expression: origins, consequences, and control.** Science 2005, **309**:2010-3.

[48]     Orian A.: **Chromatin profiling, DamID and the emerging landscape of gene expression.** Curr Opin Genet Dev 2006, **16**:157-64.

[49]     Shinar G, Dekel E, Tlusty T, Alon U.: **Rules for biological regulation based on error minimization.** Proc Natl Acad Sci USA 2006, **103**:3999-4004.

[50]     Benos PV, Bulyk ML, Stormo GD.: **Additivity in protein-DNA interactions: how good an approximation is it?** Nucleic Acids Res. 2002, **30**:4442-51.

[51]     Hollenhorst PC, Bose ME, Mielke MR, Muller U, Fox CA.: **Forkhead genes in transcriptional silencing, cell morphology and the cell cycle. Overlapping and distinct functions for FKH1 and FKH2 in Saccharomyces cerevisiae.** Genetics 2000, **154**:1533-48.

[52]    Zhu G, Spellman PT, Volpe T, Brown PO, Botstein D, Davis TN, Futcher B.:

**Two yeast forkhead genes regulate the cell cycle and pseudohyphal growth.** Nature

2000, **406:**90-4.

[53]    Wingender E, Dietze P, Karas H, Knuppel R.: **TRANSFAC: a database on**

**transcription factors and their DNA binding sites.** Nucleic Acids Res 1996, **24**:238-

41.

[54]    Sandelin A, Alkema W, Engstrom P, Wasserman WW, Lenhard B.: **JASPAR: an**

**open-access database for eukaryotic transcription factor binding profiles.** Nucleic

Acids Res 2004, **32**:D91-4.

[55]    Foat BC, Houshmandi SS, Olivas WM, Bussemaker HJ.: **Profiling condition-**

**specific, genome-wide regulation of mRNA stability in yeast.** Proc Natl Acad Sci USA

2005, **102**:17675-80.

[56]    Shannon P, Markiel A, Ozier O, Baliga NS, Wang JT, Ramage D, Amin N,

Schwikowski B, Ideker T.: **Cytoscape: a software environment for integrated models**

**of biomolecular interaction networks.** Genome Res 2003, **13:**2498-504.

[57]    Manke T, Bringas R, Vingron M.: **Correlating protein-DNA and protein-**

**protein interaction networks.** J Mol Biol 2003, **333**:75-85.

[58]    Hoffmann R, Valencia A.: **Implementing the iHOP concept for navigation of biomedical literature.** Bioinformatics 2005, **21**:252-258.

[59]    R. Kumar, D. Reynolds, A. Shevchenko, A. Shevchenko, S. Goldstone, S. Dalton.: **Forkhead transcription factors, Fkh1p and Fkh2p, collaborate with Mcm1p to control transcription required for M-phase.** Current Biology 2000, **10**:896-906.

[60]    Pic A, Lim FL, Ross SJ, Veal EA, Johnson AL, Sultan MR, West AG, Johnston LH, Sharrocks AD, Morgan BA.: **The forkhead protein Fkh2 is a component of the yeast cell cycle transcription factor SFF.** EMBO J 2000, **19:**3750-61.

# Tables

**Table (1) Functional category of 54 yeast cell cycle genes and their known transcription regulators.**

| Gene Name | Functional Category | Gene Description | Known Regulator | Literature Evidences | MIP Enrichment | $N_m/N_t$ | p-value |
|---|---|---|---|---|---|---|---|
| cln1, cln2, gic1, msb2, rsr1, bud9, mnn1, och1, psa1, gin4 , exg1, kre6, cwp1, cis3, scw4 | Budding | Genes involved in budding and in cell wall biogenesis. | Swi4, Swi6, Mbp1 | Simon et al. | 43.01.03.05 budding, cell polarity and filament formation | 7/314 | 2.34e-06 |
| | | | | | 01.05.01.01.02 polysaccharide degradation | 3/19 | 8.93e-06 |
| | | | | | 40.01 cell growth / morphogenesis | 5/190 | 4.39e-05 |
| | | | | | 42.01 cell wall | 5/215 | 7.94e-05 |
| clb5, pds5, mcm2, lrr1, cdc45, dun1, mcd1, clb6, rad51 | DNA replication & repair | Genes involved in replication, repair, and sister chromatin cohesion. | Swi4, Swi6, Mbp1 | Simon et al. | 10.03.01.01.09 G2/M transition of mitotic cell cycle | 4/50 | 3.53e-07 |
| | | | | | 10.03.04.03 chromosome condensation | 3/22 | 2.64e-06 |
| | | | | | 10.03.02 meiosis | 4/148 | 2.77e-05 |
| htb1, htb2, hta1, hta2, hho1, hhf1, hht1, tel2, hos3, arp7, ctf18 | Chromatin | Genes encoding histones, chromatin modifiers and telomere length regulators. | Mbp1, Swi6, Swi4, Fkh1 | Simon et al. | 10.01.09.05 DNA conformation modification (e.g. chromatin) | 8/185 | 4.92e-11 |
| | | | | | 16.03.01: DNA binding | 8/159 | 1.44e-11 |
| | | | | | 11.02.03.04 transcriptional control | 8/424 | 3.69e-08 |
| clb2, ace2, swi5, cdc20, apc1, tem1 | Cell cycle control | Mitosis control | Fkh1, Fkh2, Ndd1, Mcm1 | Simon et al. | 10.03.01.01.11  mitosis | 3/51 | 8.48e-06 |
| cts1, egt2 | Cytokinesis | | Ace2, Swi5, Mcm1 | Simon et al. | 10.03.03 cytokinesis (cell division) /septum formation | 2/71 | 1.13e-04 * |
| mcm3, mcm6, cdc6, cdc46 | Pre-replication complex formation | | Ace2, Swi5, Mcm1 | Simon et al. | 10.01.03.03 ori recognition and priming complex formation | 4/25 | 1.58e-10 |
| | | | | | 16.19.03 ATP binding | 4/191 | 6.72e-07 |
| | | | | | 10.01.03.01 DNA topology | 3/54 | 2.04e-06 |
| ste2, ste6, far1, mfa1, mfa2, aga1, aga2 | Mating | | Ace2, Swi5, Mcm1 | Simon et al. | 34.11.03.07 pheromone response, mating-type determination, sex-specific proteins | 7/189 | 1.39e-11 |
| | | | | | 30.05 transmembrane signal transduction | 3/27 | 2.10e-06 |

This table is based on publication of Simon et al, where 7 gene clusters are analyzed for functional enrichment of MIPS categories [28] (FunCat scheme version 2.0, March 19[th], 2004); p-values with * represents P>0.05 after correction for multiple testing; $N_t$ is the total number of yeast genes belonging to a specific MIPS category; $N_m$ is the number of genes in a specific gene cluster belonging to this category; Simon et al from [3].

**Table (2) Enrichment of MIPS functional categories (FunCat scheme version 2.0,**

**March 19[th], 2004) in 11 gene batteries.**

| Cluster Number | Gene Names | Functional Category | MIP Enrichment | $N_m/N_t$ | p-value |
|---|---|---|---|---|---|
| C_1 | Arp7, hos3, tel2, gic1, kre6, msb2 | Chromatin and Budding | 42.04 cytoskeleton | 2/114 | 4.21e-03 * |
| C_2 | Bud9, cdc45, clb5, ctf18, dun1, gin4, irr1, och1, pds5, rsr1 | Budding, DNA replication and Repair | 10.03.01.01.09 G2/M transition of mitotic cell cycle | 4/50 | 5.85e-07 |
| | | | 10.03.04.03 chromosome condensation | 3/22 | 3.77e-06 |
| C_3 | Mfa1 | Mating | 30.05 transmembrane signal transduction | 1/27 | 4.08e-03 * |
| C_4 | Apc1, cwp1, scw4, tem1 | Mitosis control and Budding | 10.03.01.01.11 mitosis | 2/51 | 3.46e-04 * |
| C_5 | ace2, cdc20, clb2, swi5 | Mitosis control | 10.03.01.01.01 G1 phase of mitotic cell cycle | 2/22 | 6.30e-05 * |
| C_6 | Cdc46, cdc6, far1, mcm2, mcm3, mcm6, ste6 | Pre-replication complex formation and Mating | 10.01.03.03 ori recognition and priming complex formation | 5/25 | 1.05e-11 |
| | | | 16.19.03 ATP binding | 6/191 | 3.65e-09 |
| | | | 10.01.03.01 DNA topology | 4/54 | 1.36e-07 |
| C_7 | Aga1, aga2, mfa2, ste2 | Mating | 34.11.03.07 pheromone response, mating-type determination, sex-specific proteins | 4/189 | 6.44e-07 |
| C_8 | clb6, cln1, cln2, mcd1, mnn1, rad51 | Budding | 18.02.01 enzymatic activity regulation / enzyme regulator | 3/78 | 3.07e-05 |
| C_9 | cis3, exg1, psa1 | Budding | 42.01 cell wall | 3/215 | 3.38e-05 |
| C_10 | cts1, egt2 | Cyokinesis | 10.03.03 cytokinesis (cell division) /septum formation | 2/71 | 1.13e-04 * |
| C_11 | hhf1, hho1, hht1, hta1, hta2, htb1, htb2 | Chromatin | 16.03.01 DNA binding | 7/159 | 4.05e-12 |
| | | | 10.01.09.05 DNA conformation modification (e.g. chromatin) | 6/185 | 3.00e-09 |
| | | | 11.02.03.04 transcriptional control | 7/424 | 4.22e-09 |

In this table, expression activities of 54 yeast genes are represented by 11 gene batteries. These gene batteries are analyzed for functional enrichment of MIPS categories [28]; p-values with * represents P>0.05 after correction for multiple testing; $N_t$ is the total number of yeast genes belonging to a specific MIPS category; $N_m$ is the number of genes in a specific gene battery belonging to this category.

**Table (3a) Results of Gaussian Graphical Models: predicted TF-gene interactions among 11 gene batteries and 9 regulators.**

| Transcriptional Factor | Regulated Cluster or Transcriptional Factor | MIP Enrichment | Literature Evidences |
|---|---|---|---|
| Fkh1 | C_1 | 42.04 cytoskeleton | Simon et al.; o |
| | C_11 | 16.03.01 DNA binding | Simon et al.; o |
| | | 10.01.09.05 DNA conformation modification (e.g. chromatin) | |
| | | 11.02.03.04 transcriptional control | |
| Fkh2 | C_1 | 42.04 cytoskeleton | Simon et al.; o |
| | C_5 | 10.03.01.01.01 G1 phase of mitotic cell cycle | Simon et al.; o |
| | C_6 | 10.01.03.03 ori recognition and priming complex formation | Simon et al.; (indirect ), o |
| | | 16.19.03 ATP binding | |
| | | 10.01.03.01 DNA topology | |
| | C_11 | 16.03.01 DNA binding | Simon et al.; o |
| | | 10.01.09.05 DNA conformation modification (e.g. chromatin) | |
| | | 11.02.03.04 transcriptional control | |
| Ndd1 | C_5 | 10.03.01.01.01 G1 phase of mitotic cell cycle | Simon et al.; o |
| | C_6 | 10.01.03.03 ori recognition and priming complex formation | Simon et al.; (indirect), o |
| | | 16.19.03 ATP binding | |
| | | 10.01.03.01 DNA topology | |
| | C_8 | 18.02.01 enzymatic activity regulation / enzyme regulator | Simon et al.; (indirect), o |
| | C_10 | 10.03.03 cytokinesis (cell division) /septum formation | Simon et al.; (indirect), o |
| Mcm1 | C_3 | 30.05 transmembrane signal transduction | Simon et al.; x |
| | C_5 | 10.03.01.01.01 G1 phase of mitotic cell cycle | Simon et al.; o |
| | C_6 | 10.01.03.03 ori recognition and priming complex formation | Simon et al.; o |
| | | 16.19.03 ATP binding | |
| | | 10.01.03.01 DNA topology | |
| | C_7 | 34.11.03.07 pheromone response, mating-type determination, sex-specific proteins | Simon et al.; o |
| Ace2 | C_4 | 10.03.01.01.11 mitosis | Simon et al.; (indirect), x |
| | C_8 | 18.02.01 enzymatic activity regulation / enzyme regulator | Simon et al.; (indirect), x |
| | C_10 | 10.03.03 cytokinesis (cell division) /septum formation | Simon et al.; o |
| | C_11 | 16.03.01 DNA binding | Simon et al.; |
| | | 10.01.09.05 DNA conformation modification | (indirect), o |
| | | (e.g. chromatin) | |
| | | 11.02.03.04 transcriptional control | |
| Swi5 | C_4 | 10.03.01.01.11 mitosis | Simon et al.; |
| | | | (indirect), o |
| | C_6 | 10.01.03.03 ori recognition and priming | Simon et al.; o |
| | | complex formation | |
| | | 16.19.03 ATP binding | |
| | | 10.01.03.01 DNA topology | |
| | C_7 | 34.11.03.07 pheromone response, mating-type determination, sex-specific proteins | Simon et al.; o |
| | C_9 | 42.01 cell wall | Simon et al.; |
| | | | (indirect), o |
| Mbp1 | C_2 | 10.03.01.01.09 G2/M transition of mitotic cell cycle | Simon et al.; o |
| | | 10.03.04.03 chromosome condensation | |
| | C_5 | 10.03.01.01.01 G1 phase of mitotic cell cycle | Simon et al.; |
| | | | (indirect), x |
| | C_6 | 10.01.03.03 ori recognition and priming | Simon et al.; |
| | | complex formation | (indirect), o |
| | | 16.19.03 ATP binding | |
| | | 10.01.03.01 DNA topology | |
| | C_8 | 18.02.01 enzymatic activity regulation / enzyme regulator | Simon et al.; o |
| Swi4 | C_7 | 34.11.03.07 pheromone response, mating-type determination, sex-specific proteins | Simon et al.; (indirect), x |
| | C_11 | 16.03.01 DNA binding | Simon et al.; o |
| | | 10.01.09.05 DNA conformation modification | |
| | | (e.g. chromatin) | |
| | | 11.02.03.04 transcriptional control | |
| Swi6 | C_4 | 10.03.01.01.11 mitosis | Simon et al.; o |
| | C_6 | 10.01.03.03 ori recognition and priming | Simon et al.; |
| | | complex formation | (indirect), o |
| | | 16.19.03 ATP binding | |
| | | 10.01.03.01 DNA topology | |
| | C_8 | 18.02.01 enzymatic activity regulation / enzyme regulator | Simon et al.; o |
| | C_10 | 10.03.03 cytokinesis (cell division) /septum formation | Simon et al.; (indirect), x |

In this table, Simon et al. means that evidence of protein-DNA interactions is available in publication [3]; indirect means that indirect evidence of protein-DNA interactions is available in publication [3]; o means that binding motif of transcription regulator is found in upstream no-coding region (web supplement [36]); x means that binding motif of transcription regulator is not found in upstream no-coding region; threshold p-value for Gaussian Graphical models is P<0.05 after correction for multiple testing.

**Table (3b) Comparison with previous work for TF-TF interactions among 11 gene batteries and 9 regulators predicted by Gaussian Graphical Models.**

| GGM Predicted TF-TF | Yu et al. Predicted TF-TF | Banerjee et al. Predicted TF-TF (PB<0.0001) | Banerjee et al. Predicted TF-TF (PB<0.001) | Banerjee et al. Predicted TF-TF (PB<0.01) | Bar-Joseph et al. Predicted TF-TF | Literature Evidences |
|---|---|---|---|---|---|---|
| Fkh2, Fkh1 | Fkh2,Fkh1 | | Fkh2, Fkh1 | Fkh2, Fkh1 | Fkh2, Fkh1 | Simon et al.; Manke et al. |
| Ndd1, Fkh1 | Ndd1, Fkh1 | | Ndd1, Fkh1 | Ndd1, Fkh1 | | Simon et al. |
| Ndd1, Fkh2 | Ndd1, Fkh2 | Ndd1, Fkh2 | Ndd1, Fkh2 | Ndd1, Fkh2 | Ndd1, Fkh2 | Simon et al.; Manke et al. |
| Mcm1, Ndd1 | Mcm1, Ndd1 | Mcm1, Ndd1 | Mcm1, Ndd1 | Mcm1, Ndd1 | Mcm1, Ndd1 | Simon et al.; Manke et al. |
| Mbp1, Fkh1 | Mbp1, Fkh1 | | | Mbp1, Fkh1 | | Simon et al. |
| Mbp1, Fkh2 | Mbp1, Fkh2 | | | | Mbp1, Fkh2 | Simon et al.; Manke et al. |
| Mbp1, Swi5 | | | | | Mbp1, Swi5 | Simon et al. |
| Swi4, Ndd1 | Swi4, Ndd1 | | | | Swi4, Ndd1 | Simon et al.; Manke et al. |
| Swi4, Mcm1 | Swi4, Mcm1 | | | | Swi4, Mcm1 | Simon et al.; Manke et al. |
| Swi4, Ace2 | Swi4, Ace2 | | | | Swi4, Ace2 | Simon et al. |
| Swi6, Fkh2 | Swi6, Fkh2 | | | Swi6, Fkh2 | | Simon et al.; Manke et al. |
| Swi6, Ndd1 | Swi6, Ndd1 | | | | | Simon et al.; Manke et al. |
| Swi6, Mbp1 | Swi6, Mbp1 | Swi6, Mbp1 | Swi6, Mbp1 | Swi6, Mbp1 | Swi6, Mbp1 | Simon et al.; Manke et al. |
| Swi6, Swi4 | Swi6, Swi4 | Swi6, Swi4 | Swi6, Swi4 | Swi6, Swi4 | Swi6, Swi4 | Simon et al.; Manke et al. |

In this table, GGM means results predicted by the Gaussian Graphical Models where threshold p-value to GGM is P<0.05 after correction for multiple testing; Yu et al. means pair of protein binding motif is frequently appeared in upstream no-coding region according to early study [8]; Banerjee et al. means results predicted by cooperativity P-value [9] and PB is P-value for TF binding to chromatin as described in Lee et al. [2];

Bar-Joseph et al. represents results predicted by GRAM modules [15]; Simon et al. from [3]; Manke et al. from [57].

**Table (4) Results of MotifSampler: predicted protein binding motifs on putative target genes of nine transcription factors.**

| Transcription Factors | Predicted Motifs (MotifSampler) | Rank order in MotifSampler | Known Binding Motifs | Literature Evidences |
|---|---|---|---|---|
| Fkh1 | TTGTTTwynT | 2 | TTGTTTACST | Yu et al. |
| Fkh2 | TTnTTTnTTT | 1 | TTGTTTACST | Yu et al. |
| Ndd1 | AGGnAAA | 1 | GTAAACA | Banerjee and Zhang. |
| Mcm1 | TTwCCynAwnrGGwAA | 1 | TTWCCCnWWWRGGAAA | Yu et al. |
| Ace2 | TACCAC | 2 | GCTGGT | Yu et al. |
| Swi5 | wGCwGC | 4 | KGCTGR | Yu et al. |
| Mbp1 | CGCGTynn | 3 | ACGCGTnA | Yu et al. |
| Swi4 | nrACGCG | 3 | TTTTCGCG | Yu et al. |
| Swi6 | nCGCGys | 1 | ACGCGT | Yu et al. |

Yu et al. means that the known binding motif is selected from publication [8]; Banerjee and Zhang mean that known binding motif is obtained from publication [9]. In this table, protein binding motif is represented by IUPAC sequence.

**Table (5) Results of Pairwise Mixed Graphical Models: predicted TF-TF interactions among 11 gene batteries and 9 regulators.**

| Transcription Factor | Regulated Transcription Factor | Literature Evidences |
|---|---|---|
| Fkh1 | Ace2 | Ihop [PMID: 10894548] |
|  | Fkh2 | Simon et al.; Pic et al.; Manke et al.; Kumar et al.; Banerjee and Zhang. |
|  | Mcm1 | Simon et al.; Kumar et al; Tsai et al. |
|  | Swi5 | Simon et al.; Ihop [PMID: 10894548] |
| Fkh2 | Mbp1 | Simon et al.; Manke et al; Tsai et al. |
|  | Mcm1 | Simon et al.; Manke et al.; Kumar et al; Tsai et al.; Banerjee and Zhang; .; Pic et al. |
|  | Ndd1 | Simon et al.; Manke et al.; Banerjee and Zhang; Tsai et al. |
|  | Swi4 | Simon et al.; Manke et al.; Tsai et al. |
|  | Swi6 | Simon et al.; Tsai et al. |
| Ndd1 | Mbp1 | Simon et al.; Manke et al. |
|  | Mcm1 | Simon et al.; Kumar et al.; Manke et al.; Tsai et al.; Banerjee and Zhang. |
|  | Swi4 | Simon et al.; Manke et al.; Tsai et al. |
|  | Swi6 | Simon et al.; |
| Ace2 | Swi5 | Simon et al.; Manke et al.; Banerjee and Zhang. |
| Mbp1 | Swi4 | Simon et al.; Manke et al. |
|  | Swi6 | Simon et al.; Manke et al.; Banerjee and Zhang; Tsai et al. |
| Mcm1 | Ace2 | Simon et al.; Pic et al. |
|  | Mbp1 | Simon et al. |
|  | Swi4 | Simon et al. |
|  | Swi5 | Simon et al. |
|  | Swi6 | Simon et al.; |
| Swi4 | Swi6 | Simon et al.; Kumar et al.; Manke et al.; Banerjee and Zhang; Tsai et al. |

In this table, Ihop means that evidence of interaction is available in the literature network [58]; Banerjee and Zhang from [9]; Kumar et al. from [59]; Manke et al. from [57]; Pic et al. from [60]; Simon et al. from [3]; Tsai et al. from [14]; threshold p-value for Pairwise Mixed Graphical Models is P<0.05 after correction for multiple testing. The full list of interactions can be found in the web supplement [36].

# Figures Legends

**Figure (1)**

Diagram of our proposed new framework for identifying combinatorial control of transcription factors: an integration of gene expression profiles, ChIP-chip data, DNA sequence information and transcription factor activities.

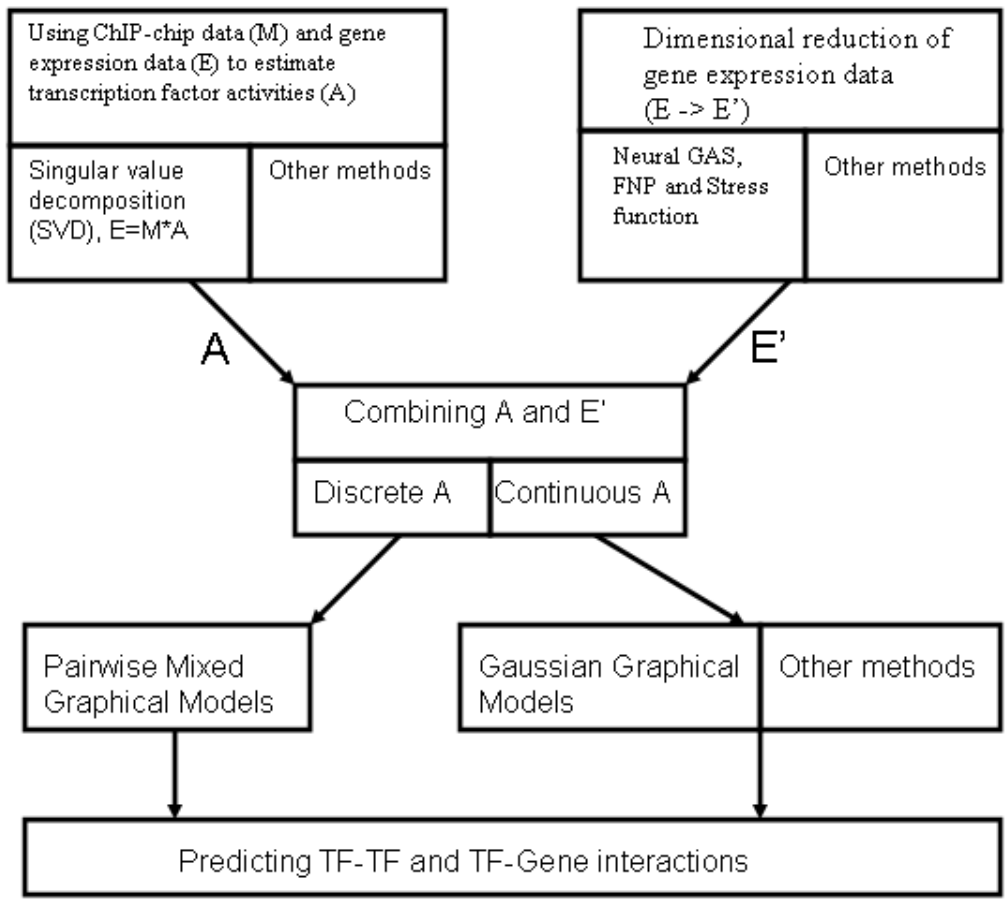

**Figure (2)**

Results of the Neural Gas algorithm and Fuzzy Nearest Prototype method that classifying 54 yeast cell cycle genes into 11 gene batteries: subplots (a-k); a red-green heat map represents log2 transferred gene expression activities (red means up regulation, green means down regulation and the color is scaled to -2 to 2); in each heat map, the last 9 columns are binding affinities (ChIP-chip data) of transcription factors and the last row is the prototype (neuron) of gene battery; in the low panel of each subplot, a red dashed line represents prototype of gene battery, blue smooth lines represent measured gene expression profiles and black smooth lines represent ChIP-chip binding data of nine transcription factors.

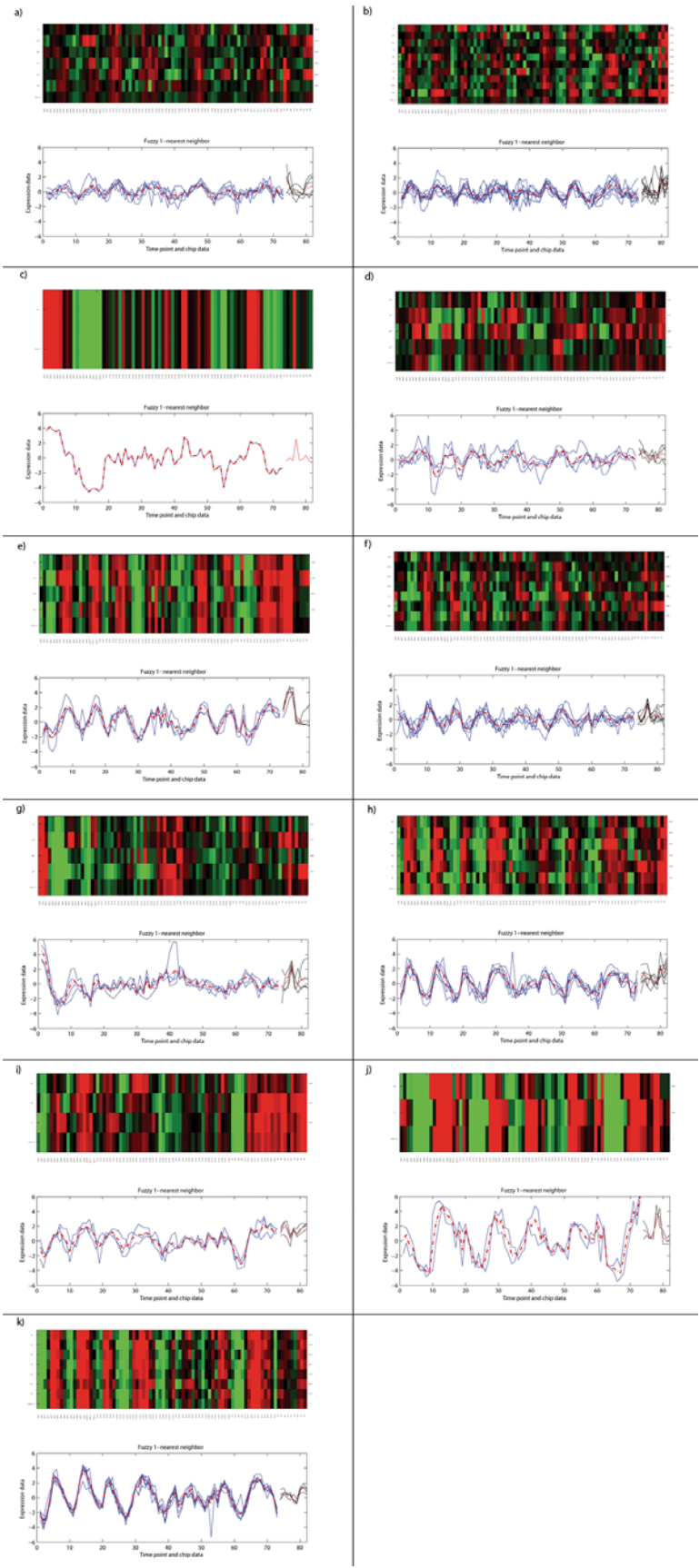

**Figure (3)**

Result of Gaussian Graphical Model with forward search algorithm: a network representation of interactions among 11 gene batteries and 9 transcription regulators in the yeast cell cycle; the gene regulatory network is visualized by Cytoscape software [56] where transcription factor is marked by the cycle and gene battery (cluster) is shown as the square; for detailed description of each gene battery please refer to Table (2).

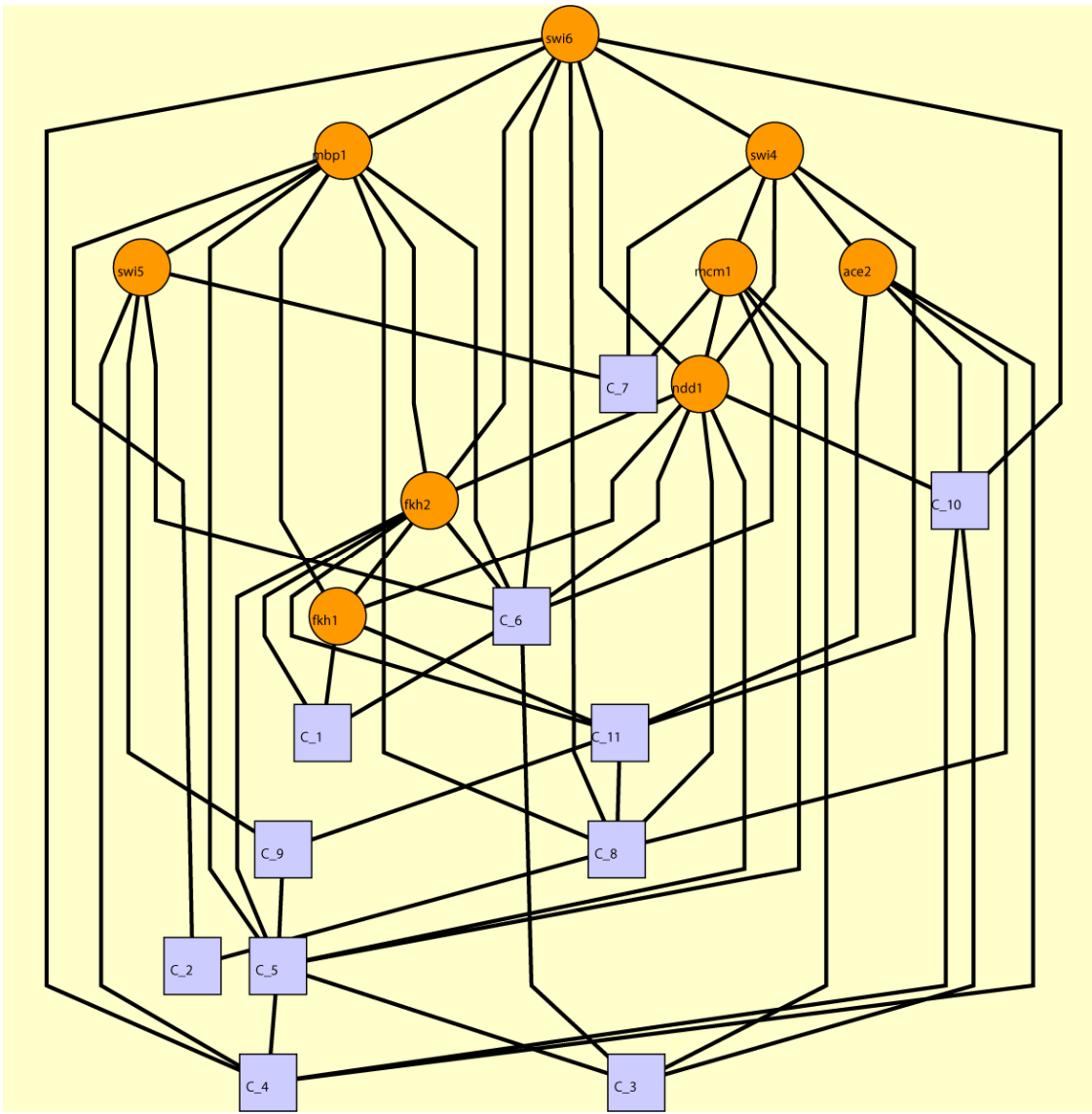

**Figure (4)**

Predicted transcription factor activation patterns of 14 pairwise TF-TF interactions: r represents correlation coefficient of pairwise TF activities and p is p-value to the correlation coefficient; in each subplot, the first TF of subtitle is marked by a red smooth line and the second TF of subtitle is marked by a black dashed line; TF-TF interactions and TF activities are computed by the Gaussian Graphical models and singular value decomposition method respectively.

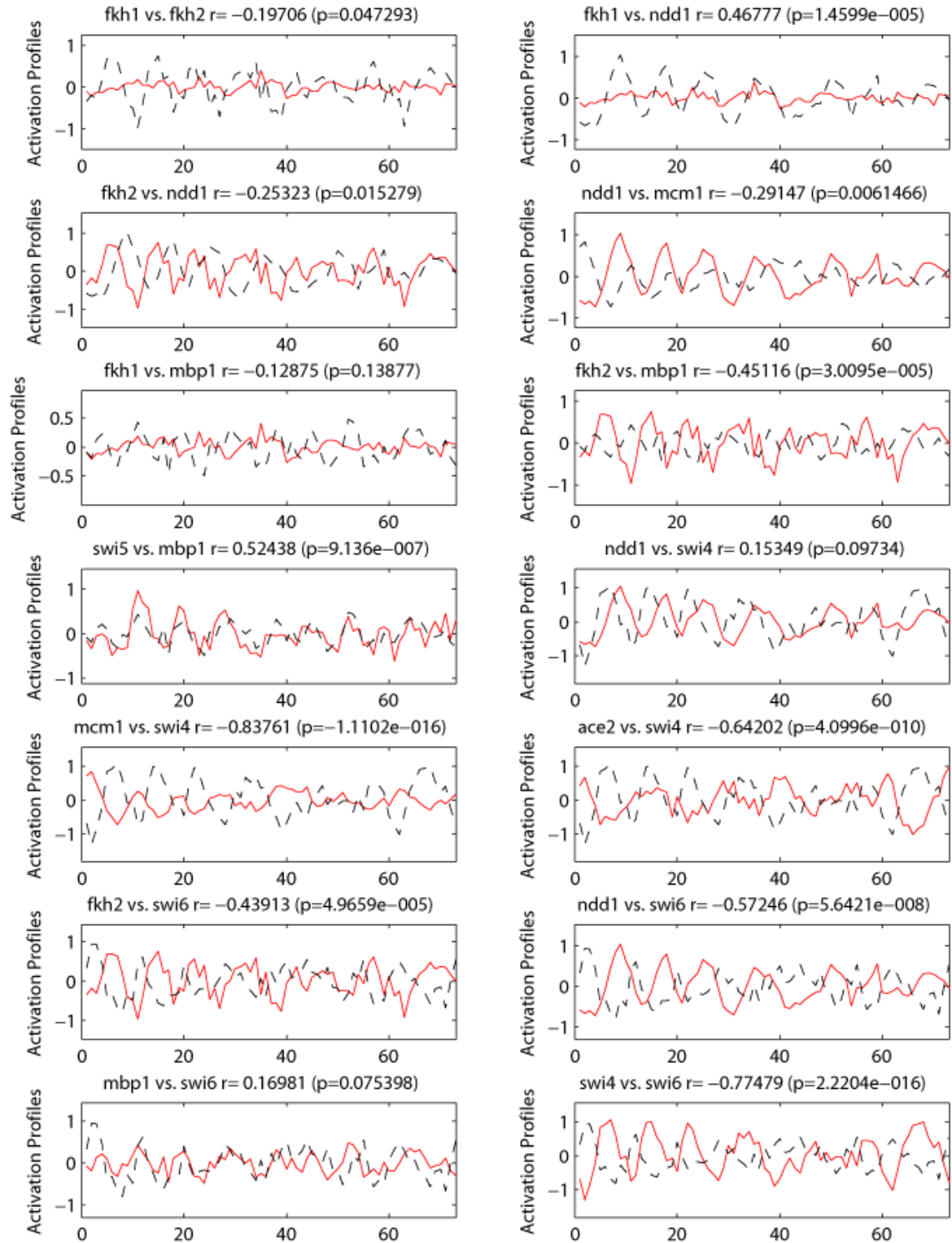

**Figure (5)**

Prototypes of 11 gene batteries: a black smooth line represents neurons W that were learned from 54 yeast genes; a red smooth line represents estimated W' that was obtained from a linear operation W'=C'A, where C' is the additive motif counts of gene batteries and A is the protein activities that was calculated by $A=M'^{-1}W$ in which M' is the additive ChIP-chip occupancy data of gene batteries; the similarity between W and W' is measured by correlation coefficient r and its p-value p.

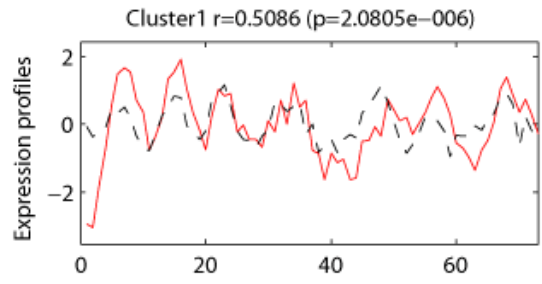
Cluster1 r=0.5086 (p=2.0805e−006)

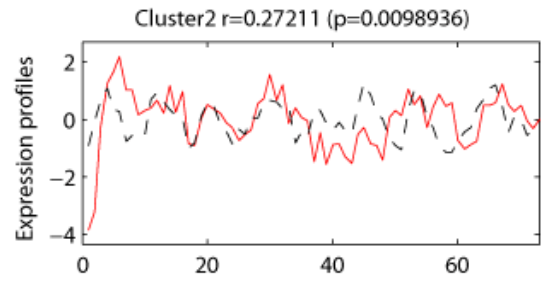
Cluster2 r=0.27211 (p=0.0098936)

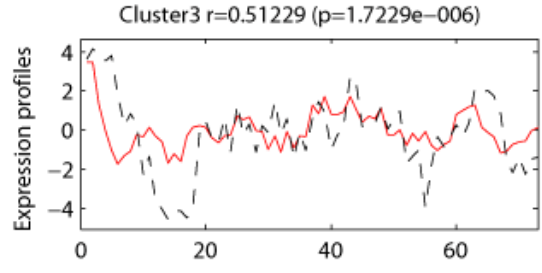
Cluster3 r=0.51229 (p=1.7229e−006)

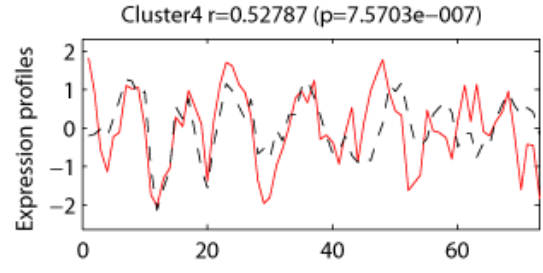
Cluster4 r=0.52787 (p=7.5703e−007)

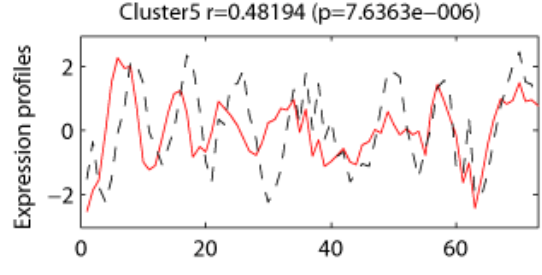
Cluster5 r=0.48194 (p=7.6363e−006)

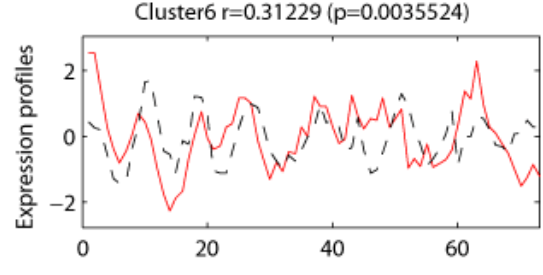
Cluster6 r=0.31229 (p=0.0035524)

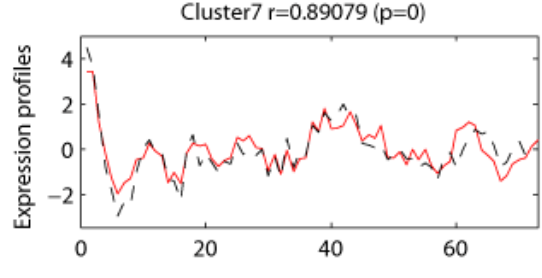
Cluster7 r=0.89079 (p=0)

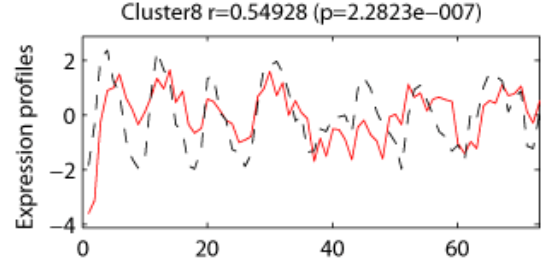
Cluster8 r=0.54928 (p=2.2823e−007)

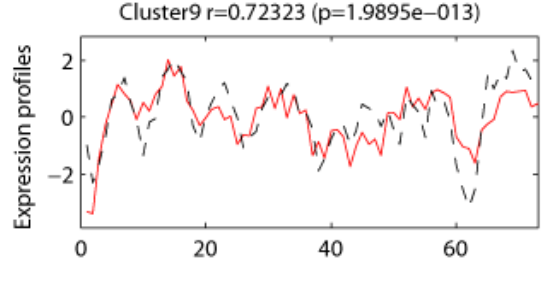
Cluster9 r=0.72323 (p=1.9895e−013)

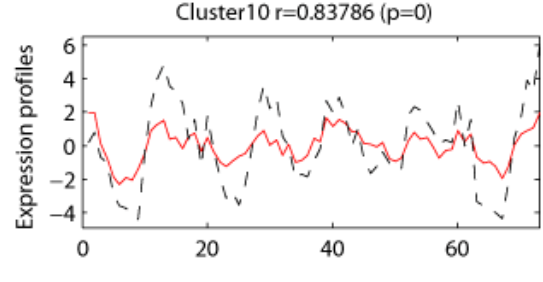
Cluster10 r=0.83786 (p=0)

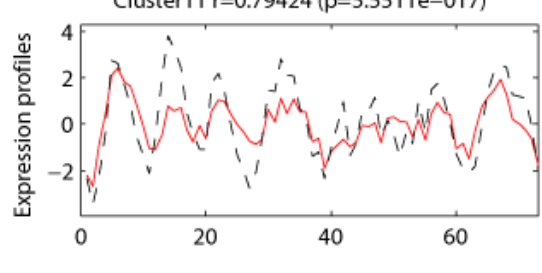
Cluster11 r=0.79424 (p=5.5511e−017)

**Figure (6)**

Using the new clustering optimization method to identify the optimal gene battery size from 54 yeast genes (the median result of ten tests):a blue smooth line with cycle is the result of 0 percent random replacement of DNA binding motif counts; a red smooth line with square is the result of 30 percent random replacement of DNA binding motif counts; a red dashed line with triangle is the result of 50 percent random replacement of DNA binding motif counts; a green smooth line with triangle is the result of 100 percent random replacement of DNA binding motif counts; red vertical line is the final result of the stress function.

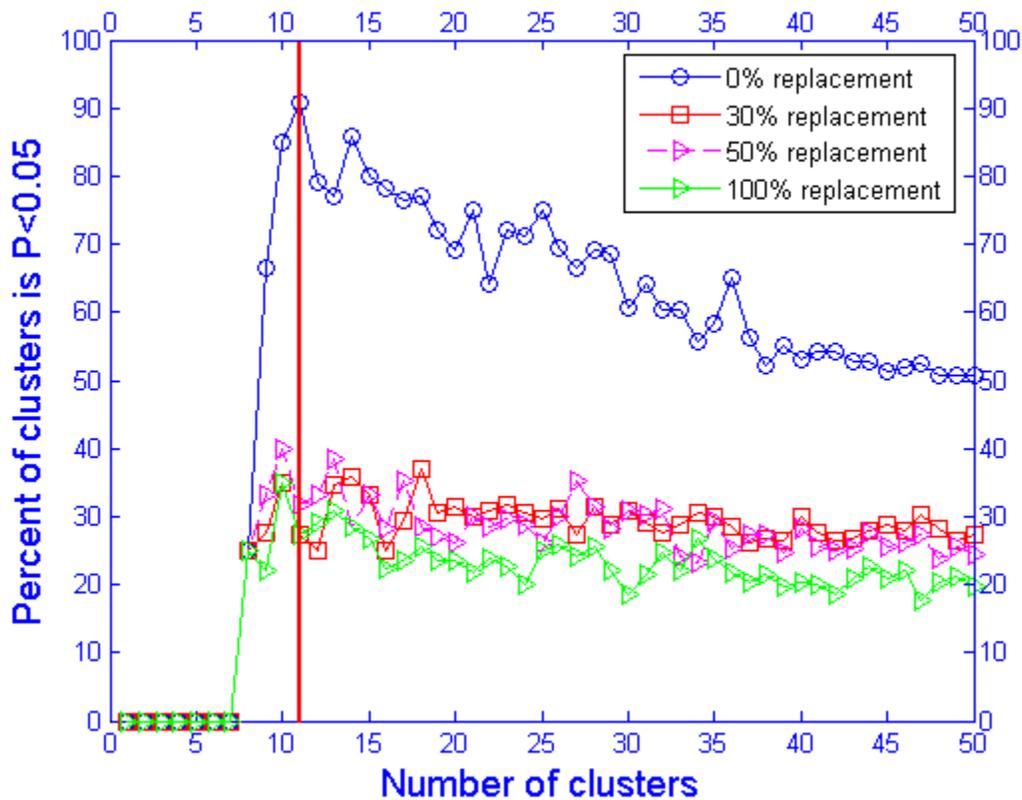

**Figure (7)**

Using the new clustering optimization method to identify the optimal gene battery size from 676 yeast genes (the median result of ten tests): a blue smooth line with square is the result of genuine DNA binding motif counts; a blue smooth line with cycle is the result of randomly sampled DNA binding motif counts; a black smooth line with cross is the outcome of stress function; a black smooth line with triangle is the Davies-Bouldin index; red vertical line is the final result of optimal subspace to 676 genes.

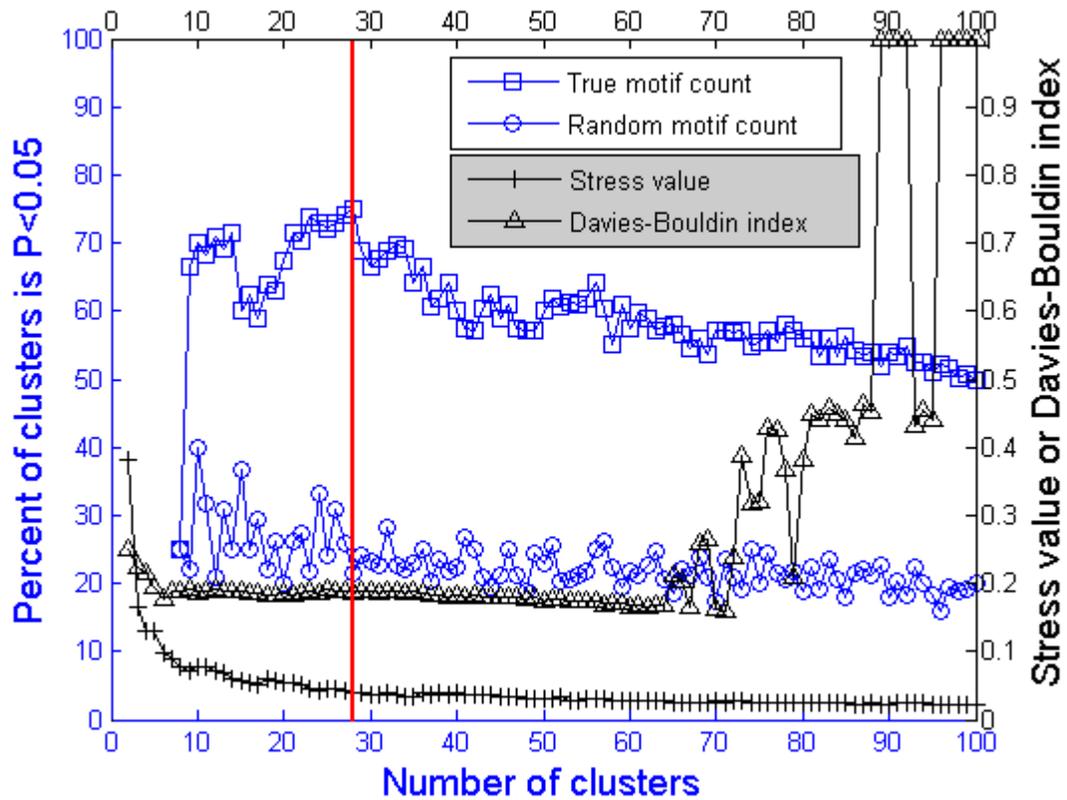